\documentclass[]{meta}

\usepackage[utf8]{inputenc}
\usepackage{array,tabularx,longtable}
\usepackage{float}
\usepackage{amsmath,amsfonts,mathtools}
\usepackage{algorithmic}
\usepackage{textcomp}
\usepackage{csquotes}
\usepackage[euler]{textgreek}
\usepackage{tablefootnote}
\usepackage{threeparttable}
\usepackage{xurl}
\usepackage{xspace}
\usepackage{enumitem}
\newlist{steps}{enumerate}{1}
\setlist[steps, 1]{label = Step \arabic*:}

\newtcolorbox{TcolorBox}[1]{fonttitle=\bfseries,title=#1}

\newlength{\imagewidth}
\usepackage{xspace}
\usepackage{dirtytalk}
\usepackage{xcolor}

\usepackage{xspace}
\usepackage{dirtytalk}
\usepackage{xcolor}

\definecolor{teal}{RGB}{0,128,128}

\newcommand{\etal}{\emph{et al.}\xspace}

\newcommand{\DefMacro}[2]{\expandafter\newcommand\csname rmk-#1\endcsname{#2}}
\newcommand{\UseMacro}[1]{\csname rmk-#1\endcsname}
\newcommand{\rom}[1]{\uppercase\expandafter{\romannumeral #1\relax}}

\begin{document}

\title{From Code Review to Code Critique: Intent, Drift, and Spotlight for AI-Generated Diffs at Scale}
\author{Chandra Maddila}
\author{Mashrur Rashik}
\author{Euna Mehnaz Khan}
\author{Smriti Jha}
\author{James Saindon}
\author{Nachiappan Nagappan}
\author[\dagger]{Peter C. Rigby}
\affiliation{Meta, USA}
\affiliation[\dagger]{Concordia University, Montreal, Canada}
\correspondence{\texttt{\{cmaddila,mrashik,eunakhan,jsaindon,nnachi,smrj\}@meta.com, pcr@meta.com}}
\abstract{AI coding agents are generating code at volumes that exceed the capacity of traditional peer review. At the same time, existing AI code review tools over-index on low-value suggestions such as style and best practices while under-indexing on the concerns human reviewers prioritize most: correctness, security, and performance. We present ARCTIC, an AI-powered Code Critique system that reframes code review around three capabilities: intent prediction, which infers why a change was made from conversation logs and metadata; drift detection, which measures divergence between the developer's intent and the agent's output via backtranslation; and code spotlight, which ranks the regions of a diff most warranting human scrutiny. We ground these capabilities in a six-theme taxonomy derived from 18,000 code reviews. Offline evaluation shows that intent prediction achieves 0.86 F1, drift detection reaches near-perfect ordinal agreement with human annotators (QWK = 0.907), and spotlight outperforms the baseline AI reviewer by 2.4x on quality estimation at 5x fewer tokens. In the experimental rollout, the drift scores reduces code misalignment by an additional 5.76 points (p = 0.026), intent prediction receives 90.2\% approval, and zero defects have been attributed to self-reviewed diffs since launch.
}
\maketitle

\section{Introduction}
\label{sec:introduction}

Code review is a a software quality assurance, serving not only as a defect-detection mechanism but also as a vehicle for knowledge transfer, standard enforcement, and team collaboration. Until recently, code review operated on a stable assumption: a manageable number of engineers produce code at a pace that peer reviewers can absorb. That assumption no longer holds. AI coding agents are generating code at unprecedented volumes and speeds, altering the practice of code review. Diffs are growing larger as agents produce multi-file changes in a single session, and external data shows that the vast majority of large code changes now ship without any formal review.

At the same time, the first generation of AI code review tools has proven insufficient. These systems tend to over-index on low-signal suggestions, such as style and best-practices comments, while under-indexing on the issues that matter most to human reviewers, namely correctness, security, and performance. Furthermore, AI coding agents frequently deviate from the developer's original intent, a phenomenon known as agentic drift, which current review tools have no mechanism to detect.

We argue that the traditional line-by-line review model is fundamentally incompatible with AI-generated code at scale. What is needed is a shift from code \emph{review} to code \emph{critique}, a system that understands diffs at a deeper level, operating simultaneously at two vantage points: the ``Forest'' view, which reasons about the high-level intent, execution milestones, and drift of a change, and the ``Trees'' view, which performs ground-level critique against security, correctness, and performance standards while directing human attention to the most consequential code regions. In this paper, we present ARCTIC, an AI-powered Code Critique system, and investigate the following research questions.

\textbf{RQ 1. Taxonomy:} \textit{What do expert human reviewers focus on during code review, and how are their concerns distributed across thematic categories?}

Understanding what constitutes a ``good'' code review comment is a prerequisite for building AI systems that can replicate or complement human review behavior. Without a principled taxonomy of review concerns, AI tools cannot be evaluated for coverage or steered toward useful critiques. We analyze over 18,000 code review comments, combining human and AI reviews that led to positive outcomes, and derive a six-theme taxonomy through iterative labeling and validation.

\textbf{RQ 2. AI Review Gap:} \textit{How do AI-generated code review comments differ from human reviewer preferences, and where do current systems over-index or under-index?}

If AI code review systems are to be trusted as a primary quality signal, we need to quantify how their output distribution compares with what human engineers actually value. Prior work has shown low adoption rates for AI suggestions, but has not mapped this gap against a structured taxonomy of review concerns. We classify 2,000 AI-generated code review comments using LLM-based tagging against our taxonomy and compute the relative difference from the human preference distribution.

\textbf{RQ 3. Intent Prediction:} \textit{Can we accurately infer the developer's intent behind a code change from available context, and which sources contribute most to prediction quality?}

Reviewers need to understand \emph{why} a change was made, not just \emph{what} changed. When AI agents produce code, the intent is often latent in conversation logs, plans, and metadata rather than explicitly stated. Accurate intent prediction is also a prerequisite for drift detection. We formalize intent prediction as a structured extraction task over multiple sources and compare agentic and zero-shot approaches on a benchmark of 121 AI-generated diffs. To the best of our knowledge, this is the first technique to extract developer intent[s] directly from coding agents' trajectories at scale.

\textbf{RQ 4. Drift Detection: } \textit{Can we reliably measure the divergence between a developer's intent and the final code produced by an AI agent?}

Agentic drift, where an AI agent deviates from the developer's goals, is an emerging risk in AI-assisted development. Recent empirical studies report that failed agent-authored pull requests commonly suffer from unwanted feature implementations and agent misalignment. Current review workflows have no mechanism to surface this divergence. We operationalize drift via backtranslation, converting the code diff into a natural-language summary and comparing it against the inferred intent, using a five-bucket rubric scored by both an LLM and human annotators on a balanced benchmark of 118 diffs.

\textbf{RQ 5. Code Spotlight:} \textit{Can an AI system reliably identify the Top-K code regions in a large diff that most warrant human scrutiny, and do these regions correlate with where real review activity occurs?}

As diffs grow larger, reviewers face an attention allocation problem: which regions should receive deep scrutiny and which are routine boilerplate? Existing AI review tools spread comments uniformly rather than prioritizing by risk. We build a Spotlight agent that produces a ranked list of regions scored by review-worthiness and evaluate on a benchmark of 298 diffs, comparing against the existing AI code reviewer.

\begin{figure*}
    \centering
    \includegraphics[width=0.9\textwidth]{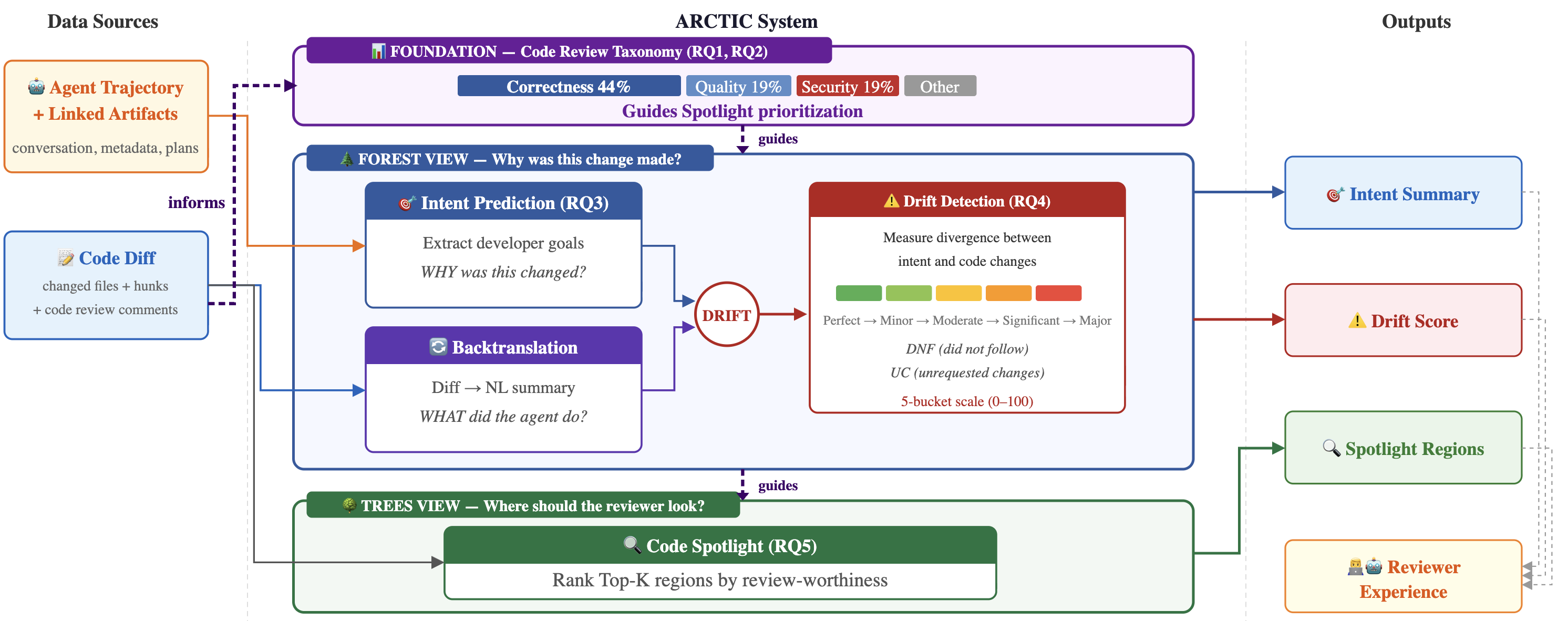}
    \caption{Overview of the ARCTIC system. Data sources (left) feed into three layers: the \textbf{Forest View} infers developer intent and detects drift via backtranslation; the \textbf{Trees View} ranks code regions by review-worthiness; and the \textbf{Foundation} provides a taxonomy-guided prioritization based on 18K human code reviews. Outputs (right) combine into a unified reviewer experience supporting self-review and directed peer review.} 
    \label{fig:arctic-overview}
\end{figure*}

Figure~\ref{fig:arctic-overview} provides an overview of the ARCTIC system architecture, showing how the five research questions feed into a unified platform. We next describe our methodology and data for each research question, followed by our offline results. We then present the system, where all components are integrated into a reimagined code review experience (Figure~\ref{fig:code-critique-ui}) that enables author self-review powered by intent, drift, and spotlight signals. We evaluate the system through a quasi-experimental drift trend study and adoption metrics from a progressive rollout. We conclude with related work, threats to validity, and directions for future research.

\section{Methodology and Data}

\subsection{Methodology for RQ 1. Taxonomy of Code Review Concerns}
\label{sec:rq1method}

To understand what expert human reviewers focus on, we derive a taxonomy of code review themes and measure the distribution of human review activity across the resulting themes.

\textbf{Data.} We use the CR2 dataset, which contains 18,000 human-reviewed diffs, each represented as a triplet of the original code, the code review comment, and the resulting code change. All entries were curated to ensure they represent actionable reviews that led to positive outcomes, specifically code changes that addressed the reviewer's feedback. We additionally collected 712 AI-generated code reviews from the current code review system, filtered to include only reviews that received positive interactions from engineers, had at least one reply comment, and where the feedback was subsequently addressed. This curated AI set represents reviews that humans found valuable enough to act on, regardless of who authored them.

\textbf{Taxonomy Creation.} We employed a four-step approach combining automated analysis with manual validation.

\begin{enumerate}
\item \textbf{Automated Theme Extraction.} We used an LLM to analyze code reviews and their corresponding diffs, identifying recurring patterns and generating candidate high-level themes and specific subcategories.

\item \textbf{Iterative Consolidation.} The LLM performed 40 iterations of pairwise comparisons to merge duplicate or overlapping themes, progressively refining the taxonomy structure. An ``Other'' catch-all category was included to capture reviews that fall outside the main classifications.

\item \textbf{Manual Validation.} Two researchers independently reviewed a subset of the generated taxonomy for relevance, coherence, and coverage, and compared the resulting themes against existing code review literature.

\item \textbf{Final Taxonomy.} The validated taxonomy consists of six high-level themes, each further subdivided into granular sub-themes.
\end{enumerate}

\textbf{Measuring the Distribution.} To measure how review activity distributes across the taxonomy, we developed a tagger using an LLM-based classification pipeline. Given a code review comment and its associated code context, the tagger assigns the review to one or more taxonomy themes in a many-to-many mapping. The pipeline has two stages. In the \emph{prompt construction} stage, we assemble a structured prompt incorporating the code review comment, its code context, and the definitions and descriptions of the six themes. In the \emph{LLM inference} stage, the prompt is processed by a large language model which generates one or more thematic tags. We treat this as a multi-class, multi-label classification problem. We applied the tagger to 2,000 randomly sampled human code reviews and the 712 curated AI reviews that were accepted by human authors to establish whether human preferences remain consistent regardless of reviewer identity.

\subsection{Methodology for RQ 2. AI Review Gap}
\label{sec:rq2method}

To quantify how AI-generated code reviews differ from human preferences, we compare the thematic distribution of AI output against the human baseline established in RQ1.

\textbf{Data.} We randomly sampled 2,000 AI-generated code reviews from the current code review system with varying outcomes, including both addressed and non-addressed reviews. Unlike the curated set in RQ1, this sample captures the full distribution of what AI systems currently produce regardless of engineer reception.

\textbf{Tagging.} We applied the same LLM-based tagger described in Section~\ref{sec:rq1method} to classify each of the 2,000 AI reviews against the six taxonomy themes.

\textbf{Gap Computation.} To quantify misalignment, we compute the relative difference for each theme:

\[
\text{Relative Difference} = \frac{\text{AI\%} - \text{Human\%}}{\text{Human\%}} \times 100
\]

A positive value indicates that AI over-represents the theme relative to human preferences, while a negative value indicates under-representation. This directly measures how far current AI review systems deviate from the priorities that human reviewers demonstrate through their actual behavior.

\subsection{Methodology for RQ 3. Intent Prediction}
\label{sec:rq3method}

To support drift detection and help reviewers understand \emph{why} a change was made, we formalize the problem of predicting developer intent from available context. An intent is not merely a summary of the code changes; a summary describes \emph{what} was changed, whereas an intent explains \emph{why} it was changed.

\textbf{Definition.} A developer intent is the underlying purpose of a code change. It answers why the change was made and captures the developer's goals, the problem they are solving, and the reasoning that led to the specific implementation. For example, while a summary might state ``Added a 5-second timeout to the external API call,'' the corresponding intent is ``prevent the application from hanging when the API becomes unresponsive.''

\textbf{Sources of Intent.} To accurately infer developer intent, we draw on multiple data sources: the developer/AI conversation, including developer messages and agent responses but excluding tool calls; change metadata, such as the title, summary, and test plan fields; AI-generated plans when the agent operates in an explicit planning mode; and linked artifacts such as project requirement documents, associated tasks, and markdown files. To avoid circular reasoning, we exclude the code change itself from the prompt when predicting intent at inference time.

\textbf{Change Gradient.} Different types of code changes require different sources and strategies for predicting intent. We identify six dimensions that affect source relevance: commit type, whether execution is human-in-the-loop or headless, the percentage of AI-authored code, who authored the change metadata, trajectory complexity, and source availability. For instance, a bug fix might rely heavily on an associated task, whereas a new feature is more likely to be grounded in a project requirement document and an AI-generated plan.

\textbf{Approaches.} We compare two intent prediction approaches. The \emph{zero-shot baseline} issues a single prompt containing the available context and instructs the LLM to extract intents in one pass. The \emph{agentic approach} uses a multi-turn loop with tool calls and iterative reasoning to explore additional context and refine the extracted intents across multiple steps. Both approaches use Opus 4.5 as the underlying model.

\textbf{Benchmark.} We constructed a benchmark of 121 AI-generated diffs, selecting for variety in diff types and AI coding assistants. For each diff, a member of the research team manually annotated ground-truth intents, informed by the diff metadata, linked artifacts, and the AI conversation trajectory.

\textbf{Evaluation.} Since developer intent is expressed in natural language, we use an LLM-as-a-Judge approach. For each predicted intent, an LLM computes pairwise semantic similarity against the ground truth. If the similarity score exceeds a threshold of 0.5, we consider the prediction a match. From these judgments we compute precision, recall, F1, Jaccard similarity, and average semantic similarity. We also report token consumption and runtime as performance metrics.

\subsection{Methodology for RQ 4. Drift Detection}
\label{sec:rq4method}

To detect when an AI agent's code changes diverge from the developer's original intent, we operationalize drift as a measurable quantity. We define drift as the divergence between the inferred intent from RQ3 and the actual code changes produced in the diff.

\textbf{Definition.} We observe two types of drift in trajectories. The first is \emph{Did Not Follow instructions} (DNF), where the agent fails to comply with the developer's explicit instructions or implied intent, for example by solving the wrong problem, ignoring stated constraints, or omitting required deliverables. DNF is by definition undesirable because the diff is not aligned with what the developer requested. The second is \emph{Unrequested Changes} (UC), where the agent makes additional changes that were not requested, even if the original task is completed correctly. UC can be beneficial, such as adding a missing null-check, or harmful, such as scope creep that increases review burden or touches sensitive files.

\textbf{Backtranslation.} To compare intent against what was actually done, we use backtranslation: converting the code diff into a natural-language summary of what changed. A backtranslation agent generates a numbered list of tasks that captures the code changes in the diff. We exclude the diff summary, test plan, and trajectory artifacts from this process because those are used in intent inference and reusing them would bias the drift computation.

\textbf{Drift Scoring Rubric.} Given the backtranslated summary and the inferred intent, the drift model, which is a zero shot LLM, evaluates divergence along two dimensions: intent satisfaction, whether each stated intent item was fully met, partially met, or unmet, and unrequested changes, whether the diff introduces out-of-scope modifications that increase review burden or introduce risk. We assign drift to one of five discrete buckets along with a continuous score on a 0-100 scale:

\begin{itemize}
\item \textbf{Perfect Alignment (0-10):} All intents fully satisfied, no bad unrequested changes.
\item \textbf{Minor Drift (11-25):} All intents satisfied with only trivial deviations.
\item \textbf{Moderate Drift (26-50):} Most intents addressed but some only partially satisfied, and/or contains noticeable bad UCs.
\item \textbf{Significant Drift (51-75):} Only some intents satisfied, and/or contains substantial bad UCs that add risk.
\item \textbf{Major Drift (76-100):} No intents satisfied, near-complete deviation from the developer's goal, and/or includes severe bad UCs like deleting functionality or introducing breaking changes.
\end{itemize}

Notably, we only penalize unrequested changes that are risky or clearly beyond scope; beneficial unrequested changes are not penalized. The same rubric is used by both the LLM and human annotators.

\textbf{Benchmark.} We constructed a balanced benchmark of 118 diffs with human-annotated ground-truth drift scores, with support across all five buckets ranging from 18 to 31 samples per bucket. Unlike a the skewed sample, this benchmark was intentionally balanced to enable meaningful per-bucket evaluation across the full drift spectrum.

\textbf{Evaluation Metrics.} We evaluate on both the ordinal and continuous scales. For ordinal evaluation, we use Quadratic Weighted Kappa (QWK) and Linear Weighted Kappa (LWK), which measure agreement between model predictions and human labels while discounting chance agreement. QWK penalizes misclassifications proportionally to the squared distance between buckets. For continuous evaluation, we report Mean Absolute Error (MAE) on the 0-100 scale. We also report per-bucket precision, recall, and F1.

\subsection{Methodology for RQ 5. Code Spotlight}
\label{sec:rq5method}

As AI-generated diffs grow in size and complexity, traditional code review can no longer keep up. Reviewers face an attention allocation problem: Which regions should receive deep scrutiny and which are routine boilerplate? AI-powered code reviews can help but existing AI reviewers over-index on low-value signals such as style and best-practices comments while under-indexing on correctness, security, and code quality, creating a signal-to-noise problem. Spotlight addresses this by surfacing only the Top-K regions of a diff that most warrant human attention, each with reasoning for why it matters.

\textbf{Approach.} Spotlight is a non-agentic, two-stage LLM reviewer: a generation pass proposes candidate regions and a critic pass validates them. Spotlight operates only on the diff and its surrounding context which include the diff title, diff summary, the modified files' contents and framework rules. The generation phase triages the diff across five code review categories derived from RQ1 and proposes candidate regions across these categories. The critic phase validates each against claim-correctness, intent, framework rules, actionability of the comment, and a senior-engineer acceptance bar, keeping only the regions that pass all five criteria. For each remaining code region, Spotlight emits a category based on the code-review taxonomy, a severity level (error, warning, or information), and an actionable rationale that contains suggested review guidance. Regions are ordered by severity and the top few are surfaced.

\textbf{Taxonomy-guided prioritization.} As noted above, Spotlight is grounded in the taxonomy from RQ1. It deliberately prioritizes the issue classes that human reviewers consistently act on, namely Correctness \& Reliability, Security, and Performance \& Efficiency, and deprioritizes the classes where current AI code review currently over-indexes, such as style and best-practices nits.

\textbf{Evaluation Setup.} We evaluate two system configurations: the Baseline AI Code Reviewer (DCR), the current AI code review system, and Spotlight, our two-stage LLM reviewer that triages a diff into classified, severity-ranked regions warranting review. Both use Opus 4.5 as the underlying model.

\textbf{Benchmark (CRBench).} We evaluate on Code Review Bench (CRBench), a code-review benchmark assembled from real landed diffs authored with AI coding assistants. Each entry is a unique diff version, identified by its diff number and version number, drawn from a rolling 60-day window of diffs. The benchmark contains roughly 300 entries and is split into two slices in a 70/30 ratio: a problematic slice of diffs that carry at least one real, reviewer-identified issue, and a clean slice of diffs whose previously flagged issues were resolved. Clean entries are constructed by taking the final version of a diff that drew actionable review comments on an earlier version but whose final version is comment-free, giving a well-matched negative class drawn from the same population as the positives rather than from arbitrary unreviewed code.

AI reviewer comments are compared against ground truth comments in the benchmark for any given diff version. Ground truth comments are filtered down to human review comments that are actionable, removing nits, questions, or praise. We additionally drop code review comments that are not determinable from the diff alone, so models are scored only on issues a reviewer could catch from the diff itself. Finally, the benchmark is stratified to match the empirical distribution of human review-comment categories such as correctness, security, and code quality so our focus remains on meaningful code review, instead of style or best practices.

\textbf{Accuracy Metrics.} We evaluate using four accuracy metrics that capture different aspects of review quality:

\begin{itemize}
\item \textbf{Quality Estimation (QE):} A triage-level judgment of whether a diff warrants a review comment at all. The system is credited for correctly flagging a diff that contains a real issue and for correctly staying silent on a clean diff, independent of whether the exact comment wording matches.
\item \textbf{Defect Localization (DL pass@5):} A rank-aware localization metric. The system passes a diff if at least one of its top 5 findings lands on the correct location of a real defect. This credits ``right place, right defect'' regardless of comment phrasing.
\item \textbf{Precision (semantic grader):} Of the findings the system emitted, the fraction that semantically match a ground-truth issue. Matching is decided by an LLM semantic grader at balanced strictness: two findings match if they describe the same underlying bug or share the same root cause. The grader groups findings by file and only compares same-file pairs.
\item \textbf{Recall (semantic grader):} Of the ground-truth issues present, the fraction that the system's findings semantically matched, under the same grader and matching rules.
\end{itemize}

\textbf{Performance Metrics.} We report two efficiency metrics: token consumption as a proxy for inference cost, and Time To First Review (TTFR), the wall-clock latency from review request to the system's first returned output.

\subsection{Methodology for Live Experiment}
\label{sec:prodmethod}

After validating each component through offline benchmarks, we progressively rolled out the ARCTIC code review workflow. ARCTIC serves as the intelligence layer powering a reimagined code review experience that enables author self-review. We describe our methodology, metrics, and progressive rollout strategy below.

\textbf{System Architecture and Funnel.} ARCTIC is a platform API that downstream surfaces consume independently. Each component, Intent, Drift, Spotlight, and Diff Categorization, is served as a separate signal. The funnel operates as follows: a diff is created, ARCTIC computes signals over it, and if the diff is eligible for self-review, the author is presented with an AI-assisted self-review experience. The author attests that the diff is correct, and the diff either lands via a streamlined path or proceeds to deferred peer review.

\textbf{Stages.} We conducted a rollout over Spring 2026:

\begin{enumerate}
\item \textbf{Intent in during code review.} Intent predictions were surfaced in the standard diff view with a thumbs-up/down feedback mechanism to collect approval signal.
\item \textbf{Drift in Diff-Next.} Drift scores were surfaced in the reimagined diff view, enabling a natural comparison between authors who used the new interface and those who did not.
\item \textbf{Bundles Early Access.} The full reimagined experience, including Intent, Drift, and Spotlight, was rolled out in an opt-in early access program for engineering teams.
\item \textbf{SDR Enablement.} The self-ship path was enabled, allowing diffs that pass ARCTIC quality signals to land without upfront peer review.
\end{enumerate}

\textbf{Goal Metrics.} We define two goal metrics for the rollout:

\begin{itemize}
\item \textbf{Intent Approval Rate:} the fraction of engineers who endorse the AI-generated intent summary via the thumbs-up/down mechanism. This measures whether the foundational signal is accurate enough for the workflow.
\item \textbf{Drift Reduction When Shown:} the additional decrease in drift score observed for authors who see the drift analysis compared to those who do not. This measures whether surfacing the signal influences author behavior.
\end{itemize}

\textbf{Adoption Metrics.} Since the self-review experience is not a randomized experiment but a progressive rollout with opt-in early access, we track adoption and conversion metrics rather than causal effect sizes:

\begin{itemize}
\item \textbf{Attestation-to-Land Rate:} of diffs where authors complete self-attestation, the fraction that successfully land via the streamlined path.
\item \textbf{Eligibility-to-Attestation Conversion:} of diffs eligible for self-review, the fraction where authors engage with and complete the self-review flow.
\end{itemize}

\textbf{Guardrail Metrics.} We monitor two guardrail metrics to ensure the new workflow does not degrade code quality:

\begin{itemize}
\item \textbf{Defect Attribution:} the number of defects attributed to diffs that went through the self-review experience.
\item \textbf{Revert Rate:} the fraction of self-reviewed diffs that are subsequently reverted, compared against the baseline revert rate for standard diffs.
\end{itemize}

\textbf{Drift Trend Study Design.} To evaluate whether ARCTIC's drift signal influences author behavior, we sampled 247 consecutive pairs of diff versions from the code review system in Spring 2026. Each pair consists of an earlier and later version of the same diff. We computed the drift score for each version and applied two filters to ensure meaningful comparisons: a minimum drift threshold (drift $>$ 25 for the earlier version, ensuring there is meaningful drift to reduce) and intent stability (LLM-rated intent similarity $>$ 0.8 between versions, ensuring the core purpose of the diff did not change). After filtering, 193 pairs remained for analysis.

We then compared the drift reduction between two groups: authors who saw the drift score in the diff-next interface (n=94) and those who did not (n=99). Group assignment was determined by whether the author used the new interface where drift is displayed, which introduces a self-selection factor that we discuss as a limitation.

\section{Results}

\subsection{RQ 1 Results. Taxonomy of Code Review Concerns}
\label{sec:rq1results}

\textit{What do expert human reviewers focus on during code review, and how are their concerns distributed across thematic categories?}

\begin{table}
\centering
\caption{RQ1 Results. The six-theme taxonomy derived from 18,000 code reviews, with the distribution of themes in human reviews and in the subset of AI reviews that human authors accepted and acted upon. The top three themes remain consistent regardless of reviewer identity.}

\begin{tabularx}{.65\textwidth}{X | r | r}
\textbf{Theme} & \textbf{Human \%} & \textbf{AI (accepted) \%} \\ \hline
Correctness \& Reliability & 44.4\% & 46.1\% \\ \hline
Code Quality \& Maintainability & 19.2\% & 19.0\% \\ \hline
Security & 19.1\% & 20.1\% \\ \hline
Best Practices \& Standards & 7.2\% & 3.5\% \\ \hline
Performance \& Efficiency & 6.6\% & 8.1\% \\ \hline
Code Design & 3.6\% & 3.3\% \\
\end{tabularx}

\label{tab:RQ1Results}
\end{table}

The iterative extraction and consolidation process produced six high-level themes, shown in Table~\ref{tab:RQ1Results}. Correctness \& Reliability covers functional soundness, bug prevention, and input validation. Code Quality \& Maintainability addresses readability, modularity, and long-term code health. Security encompasses vulnerability prevention, and access control. Best Practices \& Standards covers adherence to project conventions and documentation norms. Performance \& Efficiency concerns resource optimization. Code Design addresses high-level architectural decisions and separation of concerns.

The human distribution reveals that nearly half of all actionable code review comments concern Correctness \& Reliability (44.4\%), making it the dominant theme by a wide margin. Code Quality \& Maintainability (19.2\%) and Security (19.1\%) together account for another third. The remaining themes, Best Practices \& Standards (7.2\%), Performance \& Efficiency (6.6\%), and Code Design (3.6\%), collectively represent less than a fifth of human review activity.

Importantly, the curated subset of AI reviews that were accepted by human authors closely mirrors this distribution: Correctness \& Reliability at 46.1\%, Security at 20.1\%, and Code Quality \& Maintainability at 19.0\%. This consistency indicates that human preferences for what constitutes a valuable code review are stable regardless of whether a human or AI authored the review comment. The taxonomy captures reviewer priorities, not reviewer identity.

\begin{tcolorbox}
\textbf{RQ1 Summary.} Expert human reviewers overwhelmingly focus on Correctness \& Reliability (44.4\%), followed by Code Quality (19.2\%) and Security (19.1\%). These preferences are consistent for both human and AI-authored reviews that are acted upon by engineers.
\end{tcolorbox}

\subsection{RQ 2 Results. AI Review Gap}
\label{sec:rq2results}

\textit{How do AI-generated code review comments differ from human reviewer preferences, and where do current systems over-index or under-index?}

\begin{table}
\centering
\caption{RQ2 Results. Distribution of themes in AI-generated reviews compared to human preferences from RQ1, with relative difference quantifying over-representation (positive) or under-representation (negative). AI code review under-indexes on the themes humans value most while over-indexing on lower-priority concerns.}

\begin{tabularx}{.65\textwidth}{X | r | r | r}
\textbf{Theme} & \textbf{Human \%} & \textbf{AI \%} & \textbf{Rel. Diff.} \\ \hline
Correctness \& Reliability & 44.4\% & 25.5\% & $-$42.6\% \\ \hline
Code Quality \& Maint. & 19.2\% & 22.9\% & +19.3\% \\ \hline
Security & 19.1\% & 2.0\% & $-$89.5\% \\ \hline
Best Practices \& Standards & 7.2\% & 30.8\% & +327.8\% \\ \hline
Performance \& Efficiency & 6.6\% & 2.7\% & $-$59.1\% \\ \hline
Code Design & 3.6\% & 16.2\% & +350.0\% \\
\end{tabularx}

\label{tab:RQ2Results}
\end{table}

We applied the taxonomy-based tagger to 2,000 AI-generated code reviews sampled from the current system and computed the relative difference against the human baseline from RQ1. The results are shown in Table~\ref{tab:RQ2Results}.

The distribution of all AI-generated reviews diverges sharply from human preferences. The largest single theme in AI reviews is Best Practices \& Standards at 30.8\%, more than four times its share of human reviews, a relative difference of +327.8\%. Code Design accounts for 16.2\% of AI output compared to 3.6\% for humans, an over-representation of +350.0\%. Design reviews are often subjective, nontrivial, and require deep context about the codebase, suggesting that many of these AI suggestions may not be accurate or actionable.

Meanwhile, the themes that humans prioritize most are under-served. Security drops from 19.1\% in human reviews to just 2.0\% in AI reviews, a relative difference of $-$89.5\%. Correctness \& Reliability falls from 44.4\% to 25.5\%, a gap of $-$42.6\%. Performance \& Efficiency is under-represented by $-$59.1\%. Only Code Quality \& Maintainability shows rough parity between the two distributions, with a modest +19.3\% over-representation.

Contrasting these results with the accepted AI reviews in RQ1, we observe that the gap is a property of what AI systems currently \emph{generate}, not what humans are willing to \emph{accept}. When AI reviews do address Correctness, Security, or Code Quality, engineers act on them at comparable rates to human-authored reviews. This suggests a clear path to improvement: steering AI code review systems toward the themes that human reviewers consistently prioritize.

\begin{tcolorbox}
\textbf{RQ2 Summary.} Current AI code review under-indexes on Security  ($-$89.5\%) and Correctness \& Reliability ($-$42.6\%) while over-indexing on Code Design (+350\%) and Best Practices (+327.8\%). The gap reflects what AI generates, not what humans will accept.
\end{tcolorbox}

\subsection{RQ 3 Results. Intent Prediction}
\label{sec:rq3results}

\textit{Can we accurately infer the developer's intent behind a code change from available context, and which sources contribute most to prediction quality?}

\begin{table}
\centering
\caption{RQ3 Results. Comparison of agentic and zero-shot intent prediction across 121 AI-generated diffs. The agentic approach achieves higher F1 and precision while maintaining comparable recall, at roughly 2.3x the latency and 6.4x the token usage.}

\begin{tabularx}{.65\textwidth}{X | r | r}
\textbf{Metric} & \textbf{Agentic} & \textbf{Zero-shot} \\ \hline
Precision & 0.880 & 0.846 \\ \hline
Recall & 0.876 & 0.887 \\ \hline
F1 Score & 0.860 & 0.844 \\ \hline
Jaccard Similarity & 0.698 & 0.645 \\ \hline
Semantic Similarity & 0.772 & 0.768 \\ \hline
Tokens & 52,191 & 8,089 \\ \hline
Runtime (s) & 52.3 & 22.6 \\
\end{tabularx}

\label{tab:RQ3Results}
\end{table}

We evaluated both approaches on a benchmark of 121 AI-generated diffs and report the results in Table~\ref{tab:RQ3Results}.

\textbf{Overall quality.} The agentic approach achieves a higher F1 score than the zero-shot baseline, 0.860 versus 0.844. Unlike the earlier smaller benchmark, the precision-recall tradeoff is more nuanced at this scale: the agentic approach achieves higher precision (0.880 vs. 0.846) while the zero-shot baseline has a slight edge in recall (0.887 vs. 0.876). The agentic approach also achieves better overall set overlap as measured by Jaccard similarity (0.698 vs. 0.645), while semantic similarity scores are comparable (0.772 vs. 0.768).

\textbf{Token usage vs quality tradeoff.} The agentic approach consumes 52,191 tokens versus 8,089 for zero-shot, approximately 6.4x the token usage. Runtime is 52.3 seconds versus 22.6 seconds, roughly 2.3x the latency. This tradeoff is relevant for a system where latency and token usage must be balanced against prediction quality.

\textbf{Implications} The narrower quality gap at this larger benchmark size (F1 difference of 0.016) combined with the substantial token usage difference suggests that the zero-shot approach may be sufficient for many use cases, with the agentic approach reserved for complex diffs where its precision advantage justifies the additional tokens.

\begin{tcolorbox}
\textbf{RQ3 Summary.} On 121 diffs, the agentic intent prediction approach outperforms the zero-shot baseline (F1: 0.860 vs. 0.844) with higher precision, at 6.4x the token usage. Both approaches achieve strong absolute performance above 0.84 F1.
\end{tcolorbox}

\subsection{RQ 4 Results. Drift Detection}
\label{sec:rq4results}

\textit{Can we reliably measure the divergence between a developer's intent and the final code produced by an AI agent?}

\begin{table}
\centering
\caption{RQ4 Results. Drift detection performance on a 118-sample benchmark. The system achieves substantial agreement with human annotators (LWK=0.78) and an MAE of 10.31 on the 0-100 scale.}

\begin{tabularx}{.65\textwidth}{X | r}
\textbf{Metric} & \textbf{Value} \\ \hline
Linear Weighted Kappa (LWK) & 0.776 \\ \hline
Quadratic Weighted Kappa (QWK) & 0.907 \\ \hline
Bucket Accuracy & 64.4\% \\ \hline
MAE (0-100 scale) & 10.31 \\
\end{tabularx}

\label{tab:RQ4Results}
\end{table}

\begin{table}
\centering
\caption{RQ4 Per-Bucket Results. Precision, recall, and F1 for each drift bucket on the 118-sample benchmark. The system performs strongly on Perfect Alignment and Major Drift, with moderate performance on intermediate buckets.}

\begin{tabularx}{.65\textwidth}{X | r | r | r | r}
\textbf{Bucket} & \textbf{Prec.} & \textbf{Recall} & \textbf{F1} & \textbf{N} \\ \hline
Perfect Alignment & 0.952 & 0.833 & 0.889 & 24 \\ \hline
Minor Drift & 0.583 & 0.778 & 0.667 & 18 \\ \hline
Moderate Drift & 0.333 & 0.350 & 0.341 & 20 \\ \hline
Significant Drift & 0.474 & 0.360 & 0.409 & 25 \\ \hline
Major Drift & 0.788 & 0.839 & 0.812 & 31 \\ \hline
\textbf{Weighted Avg.} & \textbf{0.647} & \textbf{0.644} & \textbf{0.640} & \textbf{118} \\
\end{tabularx}

\label{tab:RQ4PerBucket}
\end{table}

We evaluated the drift detection system on a benchmark of 118 diffs with human-annotated ground-truth drift scores and report the results in Tables~\ref{tab:RQ4Results} and~\ref{tab:RQ4PerBucket}.

\textbf{Overall agreement.} The system achieves a Linear Weighted Kappa of 0.776 and Quadratic Weighted Kappa of 0.907, indicating substantial to near-perfect agreement between predicted drift buckets and human-annotated ground truth. The QWK of 0.907 is particularly strong, reflecting that when the system misclassifies, it typically errs by only one bucket rather than making large errors. On the continuous 0-100 scale, MAE is 10.31, indicating that predictions are typically within half a bucket boundary of the ground truth.

\textbf{Per-bucket performance.} The system performs best at the extremes of the drift spectrum. Perfect Alignment achieves F1 of 0.889 (n=24) with very high precision (0.952), and Major Drift achieves F1 of 0.812 (n=31) with strong recall (0.839). The intermediate buckets are more challenging: Minor Drift achieves F1 of 0.667 (n=18), while Moderate Drift (F1=0.341, n=20) and Significant Drift (F1=0.409, n=25) show that distinguishing between adjacent middle categories remains difficult. This is consistent with the inherently fuzzy boundary between moderate and significant divergence.

\textbf{Balanced benchmark.} This 118-sample benchmark was intentionally balanced across all five drift buckets, with support ranging from 18 to 31 per bucket. This enables meaningful per-bucket evaluation and reveals that the system's strength lies in confidently identifying well-aligned and severely drifted diffs, while the middle categories require further refinement. It is also worth noting that while the F1 score for the middle buckets are relatively low, the high LWK and QWK scores indicate that when the buckets are misclassified, they are not far off from the ground truth buckets.

\textbf{Bucket accuracy vs. agreement.} The bucket accuracy of 64.4\% may appear modest, but the high QWK (0.907) reveals that most errors are off-by-one misclassifications between adjacent buckets rather than severe disagreements. For a five-class ordinal task with inherently subjective boundaries, this represents strong performance.

\begin{tcolorbox}
\textbf{RQ4 Summary.} On 118 balanced diffs, the drift detection system achieves substantial agreement with human annotators (LWK=0.776, QWK=0.907) and MAE of 10.31. The system is strongest at identifying well-aligned and severely drifted diffs, with the middle buckets remaining more challenging.
\end{tcolorbox}

\subsection{RQ 5 Results. Code Spotlight}
\label{sec:rq5results}

\textit{Can an AI system reliably identify the Top-K code regions in a large diff that most warrant human scrutiny, and do these regions correlate with where real review activity occurs?}

\begin{table*}
\centering
\caption{RQ5 Results. Comparison of the Baseline AI Code Reviewer (AICR), which is a zero-shot prompt and Spotlight on the CRBench benchmark across problematic entries (N=208) and all entries (N=298). Spotlight substantially outperforms AICR on quality estimation and defect localization while consuming 5x fewer tokens and delivering results 6x faster.}

\begin{tabularx}{\textwidth}{X | l | r | r | r | r | r | r}
\textbf{System} & \textbf{Data set} & \textbf{QE} & \textbf{DL pass@5} & \textbf{Precision} & \textbf{Recall} & \textbf{Tokens} & \textbf{TTFR (s)} \\ \hline
Baseline (AICR) & Problematic (N=208) & 0.239 & 0.030 & 0.036 & 0.033 & 417,849 & 335.9 \\ \hline
Baseline (AICR) & All (N=298) & 0.454 & 0.048 & 0.253 & 0.255 & 417,849 & 335.9 \\ \hline
Spotlight & Problematic (N=208) & 0.582 & 0.098 & 0.108 & 0.097 & 84,757 & 55.4 \\ \hline
Spotlight & All (N=298) & 0.594 & 0.103 & 0.242 & 0.254 & 84,757 & 55.4 \\
\end{tabularx}

\label{tab:RQ5Results}
\end{table*}

We evaluated Spotlight against the Baseline AI Code Reviewer (AICR) on the CRBench benchmark, which contains 208 problematic diffs with at least one known real issue and 298 total entries including clean diffs. Results are shown in Table~\ref{tab:RQ5Results}.

\textbf{Quality Estimation.} Spotlight is far better at the triage judgment of whether a diff warrants a comment at all, improving QE by 2.4x on problematic diffs. Its edge narrows on the full set because both systems handle clean diffs well, so the gain reflects sharper discrimination of diffs that actually contain issues.

\textbf{Defect Localization.} Spotlight is 3.3x more likely than AICR to land at least one top-5 finding on the true location of a real defect, indicating its attention is directed at the
right code regions rather than spread across the diff.

\textbf{Precision and Recall.} On problematic diffs Spotlight leads AICR on both precision and recall; the two converge on the full set, where the clean diffs are scored similarly by both systems and dilute the gap. Spotlight's advantage is therefore concentrated where defect detection matters, rather than being an artifact of scoring clean code.

\textbf{Efficiency.} Spotlight delivers these gains at roughly 5x fewer tokens and 6x lower latency, bringing time-to-first-review under a minute versus nearly six for AICR. This is the
difference between interactive review feedback and a turnaround too slow to fit the workflow.

\textbf{Interpretation.} Absolute precision and recall stay modest, reflecting the inherent difficulty of predicting where human reviewers focus. Even so, consistent quality-estimation and
localization gains at a fraction of the token usage and latency show that a taxonomy-guided, region-focused reviewer beats the traditional comment-everywhere strategy for directing reviewer
attention.

\begin{tcolorbox}
\textbf{RQ5 Summary.} On the CRBench benchmark (N=298), Spotlight outperforms the baseline AI code reviewer by 2.4x on quality estimation and 3.3x on defect localization, while consuming 5x fewer tokens and delivering results 6x faster.
\end{tcolorbox}

\section{The ARCTIC System}
\label{sec:production}

After validating each component through offline benchmarks (Sections~\ref{sec:rq1results}--\ref{sec:rq5results}), we progressively rollout ARCTIC code review workflow. ARCTIC is designed as a modular platform API whose signals, Intent, Drift, Spotlight, and Diff Categorization, are served independently. This allows multiple surfaces to integrate only the primitives relevant to their use case, from the standard diff view to fully autonomous landing policies.

\subsection{ARCTIC UX}
\label{sec:produx}

\begin{figure}
    \centering
    \includegraphics[width=.80\columnwidth]{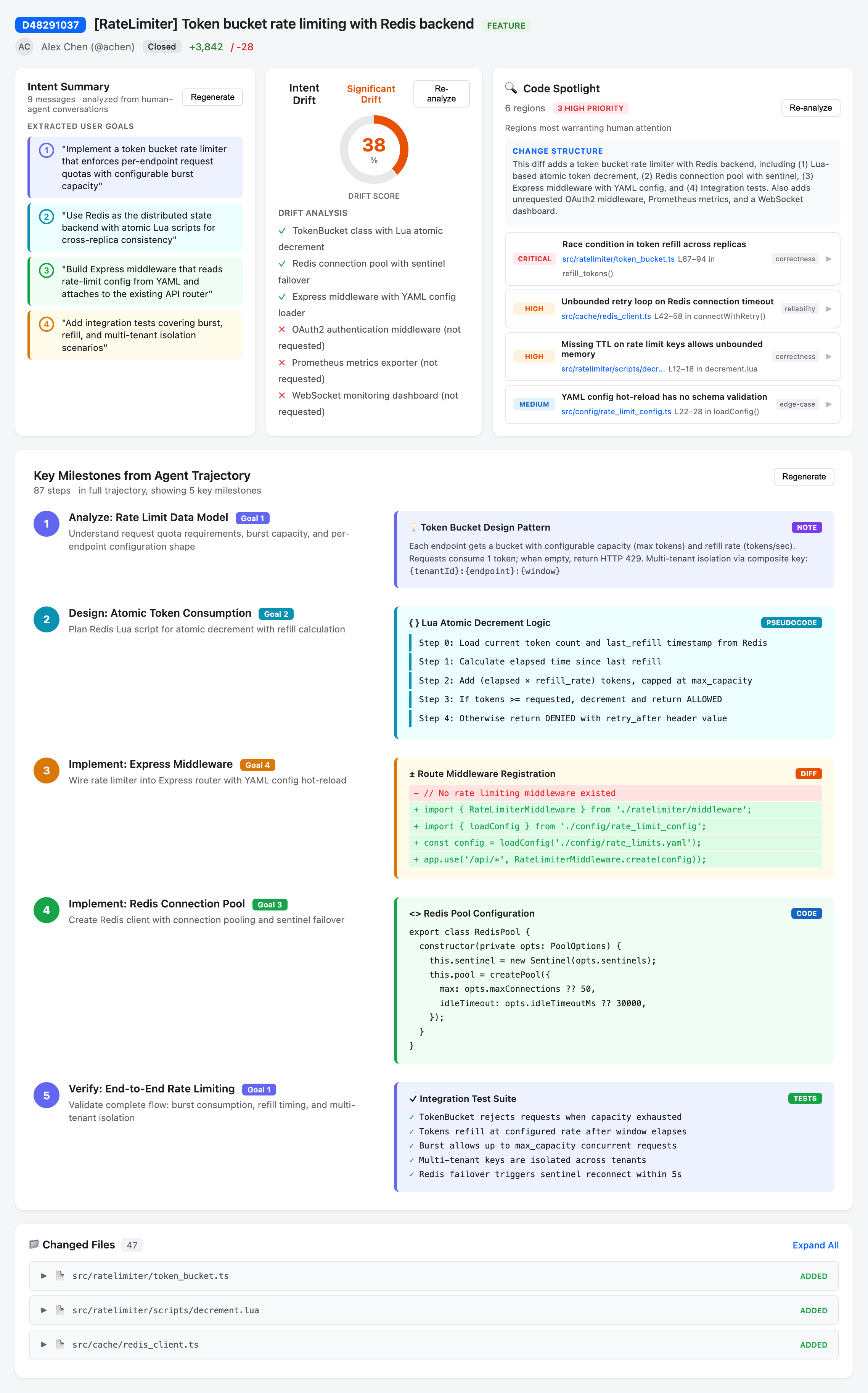}
    \caption{The ARCTIC Code Critique interface (mocked-up for a fictional diff adding retry logic, +892 lines). The top row shows \textbf{Intent Summary} (RQ3): goals extracted from the developer/AI conversation; \textbf{Intent Drift} (RQ4): a scored assessment of which goals were satisfied; and \textbf{Code Spotlight} (RQ5): the top regions ranked by review priority. Key Milestones below summarize the agent's trajectory steps.}
    \label{fig:code-critique-ui}
\end{figure}

To illustrate how ARCTIC's research contributions come together in practice, we present a mocked-up interface in Figure~\ref{fig:code-critique-ui}.

The interface is organized into two levels of abstraction. At the \emph{Forest} level, the reviewer sees the Intent Summary on the left, which surfaces the developer's goals as extracted from the human-agent conversation (RQ3). The Intent Drift panel in the center shows a continuous score with a categorical label, itemizing which goals were satisfied and which have gaps (RQ4). Together, these panels answer two questions that are otherwise invisible in a traditional diff view: \emph{what was this change trying to do?} and \emph{did the agent actually do it?}

At the \emph{Trees} level, the Code Spotlight panel ranks code regions by review priority (RQ5), categorized by concern type from the taxonomy (RQ1). Rather than reading all changed lines, the reviewer focuses on the locations most likely to contain defects. Below the top panels, Key Milestones summarize the agent's trajectory, providing evidence that the agent followed a coherent plan. The Changed Files section provides access to the traditional line-by-line view if needed.

This experience enables a new workflow where authors can self-review their diffs when quality signals indicate alignment, without requiring upfront peer review. When drift is low and Spotlight identifies no regions, authors can attest that the diff is correct and land via a streamlined path. When drift is high or Spotlight surfaces these findings, the experience directs the author to the specific gaps before routing to human reviewers. The drift score shown in this interface is the same signal evaluated in our drift trend study (Section~\ref{sec:prodresults}), where we found that authors who see it reduce drift by an additional 5.76 points compared to those who do not.

\subsection{Results from Live Experiments}
\label{sec:prodresults}

We report results from the ARCTIC rollout in Spring 2026.

\begin{table}
\centering
\caption{Production goal metrics. Intent approval reflects engineer-reported quality. The drift trend analysis demonstrates that showing drift scores to authors produces a statistically significant additional reduction in drift beyond normal iteration.}
\begin{tabularx}{.8\textwidth}{X | l | r}
\textbf{Metric} & \textbf{Design} & \textbf{Result} \\ \hline
Intent Approval Rate & Engineer feedback & 90.2\% (n=112) \\ \hline
Drift decrease across versions & Observational & $-$11.82 mean (p $<$ 0.0001) \\ \hline
Drift decrease when score shown & Quasi-experiment & $-$5.76 additional (p = 0.026) \\
\end{tabularx}
\label{tab:ProdGoal}
\end{table}

\begin{table}
\centering
\caption{Adoption and guardrail metrics from the progressive rollout of the self-review experience powered by ARCTIC signals.}
\begin{tabularx}{.65\textwidth}{X | r}
\textbf{Metric} & \textbf{Value} \\ \hline
\multicolumn{2}{l}{\textit{Adoption (weekly, as of late Spring 2026)}} \\ \hline
Attested diffs landing via streamlined path & 76\% \\ \hline
Eligibility-to-attestation conversion & 37\% \\ \hline
\multicolumn{2}{l}{\textit{Guardrail (cumulative since launch)}} \\ \hline
Defects attributed to self-reviewed diffs & 0 \\ \hline
Revert rate (self-reviewed diffs) & 0.33\% \\ \hline
Revert rate (standard diffs, baseline) & 0.17\% \\ \hline
\multicolumn{2}{l}{\textit{Scale}} \\ \hline
ARCTIC API total requests & $>$1M \\
\end{tabularx}
\label{tab:ProdAdoption}
\end{table}

\textbf{Intent Approval.} Table~\ref{tab:ProdGoal} reports that intent prediction receives a 90.2\% approval rate from engineers who interact with the predictions (101 thumbs-up, 11 thumbs-down, n=112). This indicates that the intent signal is accurate enough to serve as the foundation for the self-review workflow.

\textbf{Drift Trend Analysis.} Across 193 consecutive version pairs, drift decreased by a mean of 11.82 points between versions (one-sample t-test, t = $-$8.057, p $<$ 0.0001, Cohen's d = $-$0.58, medium effect). Among pairs whose drift changed at all, 75.9\% showed a decrease. This confirms that as developers iterate on a diff without changing their core intent, the code converges toward the intended behavior.

To test whether seeing the drift score amplifies this convergence, we compared authors who saw the drift analysis (n=94) against those who did not (n=99). Authors who saw the score reduced drift by an additional 5.76 points on average (Welch's t-test, t = $-$1.963, p = 0.026, one-sided, Cohen's d = $-$0.28, small effect). The non-parametric Mann-Whitney U test is marginally significant (p = 0.063). This provides evidence that surfacing drift scores to authors produces a measurable improvement in code alignment beyond what occurs through normal iteration.

\textbf{Adoption.} Table~\ref{tab:ProdAdoption} reports adoption metrics from the progressive rollout. Weekly authors viewing the self-review page reached 124, with 78 completing self-attestation. Of diffs that complete attestation, 76\% land via the streamlined path. The primary bottleneck is top-of-funnel conversion: only 37\% of eligible diffs complete self-attestation.

\textbf{Guardrail.} No defects have been attributed to any diff that went through the self-review experience since launch. The revert rate for self-reviewed diffs is 0.33\%, compared to a standard baseline of 0.17\%. While slightly elevated, this is within acceptable bounds for an early-access system. ARCTIC has processed over 1 million API requests across all integrated surfaces.

\begin{tcolorbox}
\textbf{Production Summary.} ARCTIC's drift signal demonstrably influences author behavior: showing drift scores produces a statistically significant additional reduction of 5.76 points (p = 0.026) beyond normal iteration. Intent prediction receives 90.2\% engineer approval, 76\% of self-attested diffs land via a streamlined path, and 0 defects have been attributed since launch.
\end{tcolorbox}

\section{Threats to Validity}
\label{sec:threats}

\subsection{Generalizability}

Our study was conducted within a single engineering environment, which may limit generalizability. The taxonomy, AI review gap, and system metrics reflect our codebase, tooling, and review culture. However, the methodology for deriving the taxonomy and the Forest/Trees decomposition are transferable to other settings.

\subsection{Construct Validity}

Several evaluations rely on LLM-as-a-Judge, including intent scoring (RQ3), drift detection (RQ4), and the semantic grader in CRBench (RQ5). While we validate against human annotations, LLM judges may exhibit systematic biases. The drift rubric's five-bucket boundaries are inherently subjective; different boundaries would yield different bucket accuracies, though the continuous MAE metric is invariant to bucketing.

\subsection{Internal Validity}

The drift trend study is quasi-experimental. Authors who used the interface where drift scores are displayed self-selected into that group, introducing potential confounds: these authors may be more engaged or working on diffs that are inherently easier to align. While the result is statistically significant (p = 0.026), we cannot rule out selection effects. Additionally, CRBench filters ground-truth comments to those determinable from the diff alone, potentially inflating precision for both systems.

\section{Related Work}
\label{sec:relatedwork}

Code review evolved from formal inspections~\cite{Fagan1976IBM} through asynchronous lightweight processes~\cite{Perry2002TSE, rigby2013convergent} to the rapid tool-supported workflows used today~\cite{sadowski2018modern}. Bacchelli and Bird~\cite{bacchelli2013expectations} found that defect detection is only one of several review outcomes, with knowledge transfer and code improvement equally important. Workload-aware reviewer recommendation~\cite{whodo2019} and nudging overdue reviews~\cite{nudge2023} further optimized the human review process.

AI-assisted code review has emerged as an active area. Vijayvergiya \etal~\cite{Vijayvergiya2024ACM} deploy AutoCommenter at Google, achieving 40\% comment resolution and over 80\% positive feedback after iterative calibration. Cihan \etal~\cite{Cihan2025SEIP} find that 73.8\% of automated comments are resolved but PR closure time increases, showing that more comments do not always add value. Li \etal~\cite{zhiyu2022automating} pre-train CodeReviewer for review generation and comment resolution. These systems generate line-level comments without distinguishing high-value from low-value findings.

Storey~\cite{Storey2026Debt} formalizes \emph{intent debt} as the absence of externalized goals that both humans and AI agents need to guide system evolution, arguing that GenAI accelerates its accumulation. Our intent prediction operationalizes detection of this debt.

Agent drift has been empirically characterized by Nanda \etal~\cite{Nanda2026FSE-industry} identify three misbehavior categories in coding agents and deploy a self-intervention system. Our drift detection complements these approaches by providing a continuous scored metric via backtranslation rather than binary classification.

Directing reviewer attention has been addressed through file-level risk scores~\cite{abreu2024drs} and comment ranking~\cite{commentfinder2022}. Our Spotlight differs by operating at the region level within files, guided by the taxonomy of concerns that human reviewers actually act on.

\section{Concluding Remarks}
\label{sec:conclusion}

AI coding agents are producing code faster than humans can review it. The traditional code review model, built on the assumption that a manageable number of engineers produce code at a pace peer reviewers can absorb, is no longer viable. We presented ARCTIC, a system that shifts code review from line-by-line inspection to structured critique organized around intent, drift, and spotlight.

Our taxonomy of 18,000 code reviews reveals that human reviewers overwhelmingly prioritize correctness and security, precisely the themes that current AI code review systems under-serve by the widest margins. This finding provides a clear direction for the field: AI review tools must be steered toward the concerns that engineers actually act on, rather than spreading attention uniformly across style, best practices, and design suggestions.

Intent prediction and backtranslation-based drift detection achieve strong agreement with human annotators, demonstrating that the divergence between what a developer asked for and what an AI agent produced can be reliably measured at scale. Spotlight outperforms the baseline AI code reviewer on both quality estimation and defect localization at a fraction of the token usage and latency, showing that a taxonomy-guided, region-focused approach to directing reviewer attention is more effective than the traditional comment-everywhere strategy.

Showing drift scores to authors produces a statistically significant reduction in code misalignment beyond what occurs through normal iteration. The self-review workflow powered by ARCTIC signals has processed over one million API requests, landed diffs at a 76\% attestation-to-land rate, and attributed zero  defects since launch. These results suggest that when developers are given the right signals about intent alignment and code risk, they can effectively self-review AI-generated changes without sacrificing quality.

As AI-generated code becomes the norm rather than the exception, we believe the review paradigm must evolve. The question is no longer ``is this code correct?'' but rather ``did the agent do what you intended, and where should you look to verify?'' ARCTIC provides a concrete instantiation of this new paradigm, and we hope it motivates further research into intent-aware, drift-sensitive, and attention-directed code critique systems.

\section*{Acknowledgments}
We would like to thank the following people for their help and support with this work: Joel Beales, Kristian Kristensen, Ale Contenti, Sherry Chen, Akshay Patel, Souvik Bhattacharya, Ilya Slavin, Chris Yoon, Victor Montalvao, Kenneth Sun, Don Stewart, Rahul Nanda, Kwaku Akoi, Stephen Kreyenbuhl, and Sid Sidhu.

Internal AI tooling and Claude Opus were used for editorial purposes. 

\section*{Data Availability}
The data in the study is bound by employee contract law and cannot be legally released even in an anonymized form.

\bibliographystyle{ACM-Reference-Format}
\bibliography{main}

\end{document}